\documentclass[a4paper, twocolumn]{article}
\usepackage{graphicx}
\usepackage[margin=1.5cm]{geometry}
\begin{document}

\def\aj{AJ}
\def\aj{AJ}  % Astronomical Journal
\def\actaa{Acta Astron.}% % Acta Astronomica
\def\araa{ARA\&A}%% Annual Review of Astron and Astrophys
\def\apj{ApJ}% % Astrophysical Journal
\def\apjl{ApJ}% % Astrophysical Journal, Letters
\def\apjs{ApJS}%% Astrophysical Journal, Supplement
\def\ao{Appl.~Opt.}%% Applied Optics
\def\apss{Ap\&SS}% % Astrophysics and Space Science
\def\aap{A\&A}%% Astronomy and Astrophysics
\def\aapr{A\&A~Rev.}% % Astronomy and Astrophysics Reviews
\def\aaps{A\&AS}%% Astronomy and Astrophysics, Supplement
\def\azh{AZh}%% Astronomicheskii Zhurnal
\def\baas{BAAS}%% Bulletin of the AAS
\def\bac{Bull. astr. Inst. Czechosl.}%% Bulletin of the Astronomical Institutes of Czechoslovakia 
\def\caa{Chinese Astron. Astrophys.}% % Chinese Astronomy and Astrophysics
\def\cjaa{Chinese J. Astron. Astrophys.}% % Chinese Journal of Astronomy and Astrophysics
\def\icarus{Icarus}%% Icarus
\def\jcap{J. Cosmology Astropart. Phys.}%% Journal of Cosmology and Astroparticle Physics
\def\jrasc{JRASC}%% Journal of the RAS of Canada
\def\mnras{MNRAS}%% Monthly Notices of the RAS
\def\memras{MmRAS}%% Memoirs of the RAS
\def\na{New A}%% New Astronomy
\def\nar{New A Rev.}%% New Astronomy Review
\def\pasa{PASA}%% Publications of the Astron. Soc. of Australia
\def\pra{Phys.~Rev.~A}% % Physical Review A: General Physics
\def\prb{Phys.~Rev.~B}%% Physical Review B: Solid State
\def\prc{Phys.~Rev.~C}%% Physical Review C
\def\prd{Phys.~Rev.~D}%% Physical Review D
\def\pre{Phys.~Rev.~E}% % Physical Review E
\def\prl{Phys.~Rev.~Lett.}%% Physical Review Letters
\def\pasp{PASP}%% Publications of the ASP
\def\pasj{PASJ}% % Publications of the ASJ
\def\qjras{QJRAS}%% Quarterly Journal of the RAS
\def\rmxaa{Rev. Mexicana Astron. Astrofis.}%% Revista Mexicana de Astronomia y Astrofisica
\def\skytel{S\&T}% % Sky and Telescope
\def\solphys{Sol.~Phys.}%% Solar Physics
\def\sovast{Soviet~Ast.}%% Soviet Astronomy
\def\ssr{Space~Sci.~Rev.}% % Space Science Reviews
\def\zap{ZAp}%% Zeitschrift fuer Astrophysik
\def\nat{Nature}%% Nature
\def\iaucirc{IAU~Circ.}%% IAU Cirulars
\def\aplett{Astrophys.~Lett.}%% Astrophysics Letters
\def\apspr{Astrophys.~Space~Phys.~Res.}%% Astrophysics Space Physics Research
\def\bain{Bull.~Astron.~Inst.~Netherlands}%% Bulletin Astronomical Institute of the Netherlands
\def\fcp{Fund.~Cosmic~Phys.}%% Fundamental Cosmic Physics
\def\gca{Geochim.~Cosmochim.~Acta}%% Geochimica Cosmochimica Acta
\def\grl{Geophys.~Res.~Lett.}%% Geophysics Research Letters
\def\jcp{J.~Chem.~Phys.}% % Journal of Chemical Physics
\def\jgr{J.~Geophys.~Res.}% % Journal of Geophysics Research
\def\jqsrt{J.~Quant.~Spec.~Radiat.~Transf.}% Journal of Quantitiative Spectroscopy and Radiative Trasfer
\def\memsai{Mem.~Soc.~Astron.~Italiana}%% Mem. Societa Astronomica Italiana
\def\nphysa{Nucl.~Phys.~A}% % Nuclear Physics A
\def\physrep{Phys.~Rep.}% Physics Reports
\def\physscr{Phys.~Scr}%% Physica Scripta
\def\physrev{Phys.~Rev.}% Physics Reports
\def\planss{Planet.~Space~Sci.}%% Planetary Space Science
\def\procspie{Proc.~SPIE}
\def\amp{\&}
\def\euro{\mbox{\raisebox{.25ex}{{\it =}}\hspace{-.5em}{\sf C}}}
% Bibliography and bibfile
\def\aj{AJ}%
          % Astronomical Journal
\def\actaa{Acta Astron.}%
          % Acta Astronomica
\def\araa{ARA\&A}%
          % Annual Review of Astron and Astrophys
\def\apj{ApJ}%
          % Astrophysical Journal
\def\apjl{ApJ}%
          % Astrophysical Journal, Letters
\def\apjs{ApJS}%
          % Astrophysical Journal, Supplement
\def\ao{Appl.~Opt.}%
          % Applied Optics
\def\apss{Ap\&SS}%
          % Astrophysics and Space Science
\def\aap{A\&A}%
          % Astronomy and Astrophysics
\def\aapr{A\&A~Rev.}%
          % Astronomy and Astrophysics Reviews
\def\aaps{A\&AS}%
          % Astronomy and Astrophysics, Supplement
\def\azh{AZh}%
          % Astronomicheskii Zhurnal
\def\baas{BAAS}%
          % Bulletin of the AAS
\def\bac{Bull. astr. Inst. Czechosl.}%
          % Bulletin of the Astronomical Institutes of Czechoslovakia 
\def\caa{Chinese Astron. Astrophys.}%
          % Chinese Astronomy and Astrophysics
\def\cjaa{Chinese J. Astron. Astrophys.}%
          % Chinese Journal of Astronomy and Astrophysics
\def\icarus{Icarus}%
          % Icarus
\def\jcap{J. Cosmology Astropart. Phys.}%
          % Journal of Cosmology and Astroparticle Physics
\def\jrasc{JRASC}%
          % Journal of the RAS of Canada
\def\mnras{MNRAS}%
          % Monthly Notices of the RAS
\def\memras{MmRAS}%
          % Memoirs of the RAS
\def\na{New A}%
          % New Astronomy
\def\nar{New A Rev.}%
          % New Astronomy Review
\def\pasa{PASA}%
          % Publications of the Astron. Soc. of Australia
\def\pra{Phys.~Rev.~A}%
          % Physical Review A: General Physics
\def\prb{Phys.~Rev.~B}%
          % Physical Review B: Solid State
\def\prc{Phys.~Rev.~C}%
          % Physical Review C
\def\prd{Phys.~Rev.~D}%
          % Physical Review D
\def\pre{Phys.~Rev.~E}%
          % Physical Review E
\def\prl{Phys.~Rev.~Lett.}%
          % Physical Review Letters
\def\pasp{PASP}%
          % Publications of the ASP
\def\pasj{PASJ}%
          % Publications of the ASJ
\def\qjras{QJRAS}%
          % Quarterly Journal of the RAS
\def\rmxaa{Rev. Mexicana Astron. Astrofis.}%
          % Revista Mexicana de Astronomia y Astrofisica
\def\skytel{S\&T}%
          % Sky and Telescope
\def\solphys{Sol.~Phys.}%
          % Solar Physics
\def\sovast{Soviet~Ast.}%
          % Soviet Astronomy
\def\ssr{Space~Sci.~Rev.}%
          % Space Science Reviews
\def\zap{ZAp}%
          % Zeitschrift fuer Astrophysik
\def\nat{Nature}%
          % Nature
\def\iaucirc{IAU~Circ.}%
          % IAU Cirulars
\def\aplett{Astrophys.~Lett.}%
          % Astrophysics Letters
\def\apspr{Astrophys.~Space~Phys.~Res.}%
          % Astrophysics Space Physics Research
\def\bain{Bull.~Astron.~Inst.~Netherlands}%
          % Bulletin Astronomical Institute of the Netherlands
\def\fcp{Fund.~Cosmic~Phys.}%
          % Fundamental Cosmic Physics
\def\gca{Geochim.~Cosmochim.~Acta}%
          % Geochimica Cosmochimica Acta
\def\grl{Geophys.~Res.~Lett.}%
          % Geophysics Research Letters
\def\jcp{J.~Chem.~Phys.}%
          % Journal of Chemical Physics
\def\jgr{J.~Geophys.~Res.}%
          % Journal of Geophysics Research
\def\jqsrt{J.~Quant.~Spec.~Radiat.~Transf.}%
          % Journal of Quantitiative Spectroscopy and Radiative Trasfer
\def\memsai{Mem.~Soc.~Astron.~Italiana}%
          % Mem. Societa Astronomica Italiana
\def\nphysa{Nucl.~Phys.~A}%
          % Nuclear Physics A
\def\physrep{Phys.~Rep.}%
          % Physics Reports
\def\physscr{Phys.~Scr}%
          % Physica Scripta
\def\planss{Planet.~Space~Sci.}%
          % Planetary Space Science
\def\procspie{Proc.~SPIE}%
          % Proceedings of the SPIE
%\title{Introduction to \LaTeX{}}
%\author{Author's Name}
\title{Constraining multiple systems with GAIA}
\author{Beauvalet, L.$^1$, Lainey, V.$^1$, Arlot, J.-E.$^1$, Bancelin, D.$^1$, Binzel, R. P.$^{2,1}$, Marchis, F.$^{3,1}$}

\maketitle
\noindent 1. IMCCE-Observatoire de Paris-UPMC, 77, avenue Denfert-Rochereau, 75014, Paris, France\\
2. Department of Earth, Atmospheric, and Planetary Sciences, Massachusetts Institute of Technology, 77 Massachusetts Avenue, Cambridge, MA 02139 USA\\
3. SETI Institute, 515 N. Whisman Road, Mountain View, CA 94043, USA
\begin{abstract}
GAIA will provide observations of some multiple asteroid and dwarf systems. These observations are a way to determine and improve the quantification of dynamical parameters, such as the masses and the gravity fields, in these multiple systems. Here we investigate this problem in the cases of Pluto's and Eugenia's system. We simulate observations reproducing an approximate planning of the GAIA observations for both systems, as well as the New Horizons observations of Pluto. We have developed a numerical model reproducing the specific behavior of multiple asteroid system around the Sun and fit it to the simulated observations using least-square method, giving the uncertainties on the fitted parameters. We found that GAIA will improve significantly the precision of Pluto's and Charon's mass, as well as Petit Prince's orbital elements and Eugenia's polar oblateness.
\end{abstract}

\section{Introduction}
Astrometric monitoring of multiple systems is a powerful way in the Solar System to have access to the physical properties of small bodies. Indeed, the satellite motions provide the mass of the primary, as well as the harmonics of its gravity field. The bodies involved can be very different in size and masses, from dwarf planets to small asteroids. These systems can be very compact and, as a result, are difficult to observe from Earth without adaptive optics. We can expect GAIA to observe the components of such systems, both the primary and its satellites, if these later are far enough from the primary \cite{Bancelin}. GAIA will make precise and regular observations of them, and as a result, will probably improve our knowledge of their dynamical parameters. 

The purpose of this paper is to estimate the precision on the  dynamical parameters we can expect for multiple systems thanks to GAIA. This kind of systems have been discovered among nearly every family of small bodies. 
We have investigated the contribution of GAIA's observations for one system in the Kuiper Belt : Pluto, and one in the Main Belt : 45 Eugenia. After a presentation of the dynamical model used especially to describe multiple systems in Section 1, we will develop the case of each system in a different section.

\section{Dynamical model}
We use here the same numerical model which has been developed in \cite{Moi}. We consider the motion of every bodies of a multiple system in the inertial reference frame ICRF centered on the barycenter of the Solar System. We compute the motion of the bodies disturbed by the Sun and the planets, whose positions are obtained through the numerical ephemeris DE405 \cite{Standish1998}. The initial positions, velocities and masses in Pluto's system  come from \cite{Tholen2008}. When needed, we include the second order harmonics of the polar oblateness of the primary, $J_2$. We use the following notations :
\begin{itemize}
\item $i$  an integrated body from the considered system, 
\item$j$ the Sun or a planet, 
\item$m_{i}$ the mass of the body $i$, 
\item${\bf{r}}_j$ the position vector of the body $j$ with respect to the Solar System barycenter, 
\item$r_{ij}$ the distance between bodies $i$ and $j$, 
\item$R_l$ the equatorial radius of body $l$, $J_2^{(l)}$ the polar oblateness of body $l$,
\item $U_{\bar{i}\, \hat{l}}$ the potential of the $l$ body's oblateness on the $i$ body's center of mass. 
\end{itemize}

We then obtain the following equation of motion :
\begin{equation}\label{eqmotion}
\begin{array}{r l}
 \ddot{\bf{r}}_i=& \displaystyle\sum_{j=1}^{10} -\frac{Gm_j(\bf{r}_i-\bf{r}_j)}{r_{ij}^3} \\
&\displaystyle + \sum_{l=1,\ l\ne i}^{ 3} \Bigl(-\frac{Gm_l(\bf{r}_i-\bf{r}_l)}{r_{il}^3}  \\
& \displaystyle + Gm_l \nabla_lU_{\bar{l}\, \hat{i}} - Gm_l\nabla_i U_{\bar{i}\, \hat{l}}\Bigr)
 \end{array}
\end{equation}
where $U_{\bar{i}\, \hat{l}}$ is a function of the oblateness of $l$, and $\phi_i$ the latitude of $i$ with respect to $l$ equator :
\begin{equation}
U_{\bar{i}\, \hat{l}}= -\frac{R_l^2 }{r_{il}^3} J_{2}^{(l)}\left(\frac{3}{2}\sin^2 (\phi_i) -\frac{1}{2}\right) 
\end{equation}
Numerical integration of the equations of motion has been made using the 15th order Gauss-Radau integrator developed by \cite{Everhart1985}.
Then we can adjust our model to the observations through a least-square procedure without constraints. 

\section{Pluto's sytem}
\subsection{Description}

In 2013, Pluto will be about 33 AU from the Sun. We consider a four bodies system : Pluto, its most massive satellite Charon \cite{chardisc}, Nix and Hydra \cite{Weaver2005}. Physical details on these objects are given in Table \ref{Plusyst}. A last satellite has been discovered in July 2011 \cite{2011P4} and has not been included in the model.

\begin{table}

\begin{tabular}{ccc}
\hline\hline &Pluto&Charon\\
\hline
semi-major axis & 39.26 AU &  19570.45 km \\
diameter& 2340 km & 1206 km\\
angular diameter& 100 mas & 55 mas \\
magnitude & 15.1 & 16.8 \\
GM (km$^3$ s$^{-2}$)&870.3 & 101.4\\
\hline
&Nix&Hydra\\
\hline
semi-major axis & 49242. km & 65082. km\\
diameter& 88 km & 72 km\\
angular diameter& 4 mas & 3 mas\\
magnitude & 23.7 & 23.3\\
GM (km$^3$ s$^{-2}$)&0.039 & 0.021\\
\hline
\end{tabular}\caption{Characteristics of Pluto and its satellites}\label{Plusyst}
\end{table}
The particularity of Pluto among other systems in the Kuiper Belt is that it will be observed \textit{in situ} by the probe New Horizons in 2015. This means that the system will then be observed simultaneously by this probe and by GAIA. Nonetheless, New Horizons will observe the four bodies of the system during a very short amount of time, whereas GAIA will only be able to detect Pluto and Charon, but its observations will be regularly made during five years. 
The contribution of New Horizons' observations to our knowledge of the dynamical parameters of the system has been investigated in \cite{Moi}. In this previous study, we have found that the only parameters which can be estimated are the masses of the bodies, while the oblate gravity fields will not be obtained, even at the time of New Horizons arrival. We use here the same method for GAIA.

\subsection{Data simulation}
Our goal is to determine the precision on the masses we can expect thanks to GAIA's observations. To do so, we simulate data at the moment of already existing and expected future observations of the system. We then fit our model to the simulations and extract the 1-$\sigma$ uncertainty from the least-square method. We did not include noise to our simulation. This comes from the fact that we only want the statistical uncertainty, a quantity which depends only on the uncertainty of the observations, the influence of the parameter on the system and the correlations between the parameters.
 We used the Rendez-vous software to obtain a possible schedule of observations of Pluto's sytem by GAIA between 2013 and 2017. This software code has been developed by C. Ordenovic, F. Mignard and  P. Tanga (OCA) for Gaia DPAC. The uncertainty attached to these simulated GAIA observations is considered to be 1 mas. 
 As a first approximation, we neglect the fact that this precision is available only in the direction of the scan.
 
The dates used for GAIA simulations are given in Table \ref{obsplugaia}. The simulated ground-based observations consist of ten observations per year, with the same uncertainties as current ground-based observations. The New Horizons simulations are obtained using a preliminary schedule of the mission, as well as its estimated uncertainties. More details on these two sets of simulations are given in \cite{Moi}.
 \begin{table}
\caption{Dates used for the simulation of GAIA observations for Pluto's system.}\label{obsplugaia}
\begin{center}
\begin{tabular}{cc}
\hline\hline
Dates  \\
\hline
29/04/2013 17:16:31  & 05/04/2014 20:14:44 \\
29/04/2013 19:03:05  & 23/08/2014 07:18:58 \\
13/05/2013 13:08:54  & 23/08/2014 13:19:11 \\
13/05/2013 17:22:33  & 23/08/2014 09:05:32 \\
13/05/2013 19:09:06  & 07/09/2014 07:24:37 \\
16/08/2013 14:40:26  & 07/09/2014 09:11:11 \\
16/08/2013 18:54:05  & 11/10/2014 18:49:60 \\
16/08/2013 20:40:39  & 11/10/2014 20:36:33 \\
17/08/2013 00:54:18  & 25/02/2015 13:29:17 \\
17/08/2013 02:40:52  & 25/02/2015 15:15:51 \\
17/08/2013 06:54:31  & 15/03/2015 15:19:11 \\
17/08/2013 08:41:04  & 17/04/2015 01:03:56 \\
20/08/2013 18:57:21  & 01/09/2015 15:35:38 \\
20/08/2013 20:43:54  & 22/09/2015 07:48:31 \\
21/08/2013 00:57:33  & 22/09/2015 09:35:05 \\
21/08/2013 02:44:07  & 22/10/2015 15:11:01 \\
21/08/2013 06:57:46  & 06/03/2016 14:02:18 \\
21/08/2013 08:44:20  & 29/03/2016 15:43:50 \\
30/09/2013 18:15:16  & 27/04/2016 01:39:15 \\
30/09/2013 20:01:50  & 10/09/2016 20:21:33 \\
07/11/2013 19:22:56  & 10/09/2016 22:08:07 \\
07/11/2013 23:36:35  & 05/10/2016 21:58:12 \\
08/11/2013 01:23:09  & 01/11/2016 07:58:37 \\
16/11/2013 11:42:28  & 01/11/2016 09:45:11 \\
16/11/2013 13:29:02  & 17/03/2017 04:23:16 \\
16/11/2013 17:42:41  & 13/04/2017 10:08:01 \\
16/11/2013 19:29:15  & 07/05/2017 16:00:14 \\
17/02/2014 02:45:53  & 21/09/2017 08:55:12 \\
17/02/2014 06:59:32  & 21/09/2017 10:41:46 \\
17/02/2014 08:46:06  & 20/10/2017 10:21:17 \\
27/02/2014 13:06:04  & 11/11/2017 14:31:42 \\
27/02/2014 14:52:38  & 11/11/2017 16:18:16 \\
05/04/2014 18:28:10  &  	           \\
\hline
\end{tabular}
\end{center}
\end{table}
 
 As a result, we have two different sets of simulated data :
 \begin{itemize}
 \item 1992-2014+NH : reproducing the existing observations, the future possible observations and New Horizons temporary observation schedule
 \item GAIA : reproducing observations of the system by GAIA
 \end{itemize}
Simulations with GIBIS suggest that the two bodies, Pluto and Charon, should always be detected as separated objects. Nix and Hydra have respective magnitude of about 23.7 and 23 \cite{2006astro.ph..5014S}, so these two bodies will not be detected by GAIA.

\subsection{Results}
\subsubsection{Contribution of GAIA's whole mission}
The uncertainties on the masses for every object of the system is given in Table \ref{Res1Plu}. We can see that GAIA's observations will lower the uncertainties of the masses of Pluto and Charon when comparing the cases with and without GAIA. We can also see that the precision of the masses of Nix and Hydra will also be a little bit modified. This comes from the fact that, if a parameter is more constrained, the fitting process can no longer reduce the residuals through changing this parameter as much as before. As a result, the role, and so the uncertainty, of the other parameter will also change.
The masses which will be the most improved by GAIA are Pluto's and Charon's. We could have expected that the uncertainties on the masses of Nix and Hydra would have been much lowered because of the stronger constraints put on Pluto and Charon. This is not what happens here, only the uncertainty on Hydra's mass being lowered. This comes from the fact that the correlations between the parameters do not decrease linearly with the number of observations available.

%One of the more puzzling consequence of the GAIA's observations is that they will not improve only the masses of Pluto and Charon, but those of Nix and Hydra as well. This can be explained by the fact that the bodies which have the most influence on these objects are Pluto and Charon. When fitting the orbit, if the parameters with a major influence are more constrained, the residuals can no longer be reduced by fitting this parameter. Hence, constraining Pluto's and Charon's dynamical parameters means that those of Nix and Hydra will have a clearer effect, and hence an expected higher precision.

\subsubsection{Orbit enhancement before New Horizons arrival}
GAIA will be launched two years before New Horizons' arrival, and so will collect data before the fly-by. We have searched whether GAIA would enable us to put stronger constraints on the masses and, as a result, on the body ephemerides. We have used the same method as before with simulations spanning between 2013 and 2015 before New Horizons' fly-by. The set of simulations named 1992-2015 uses the dates of the currently available observations, the future observations before New Horizons' arrival and the observations from GAIA before 2015.

The obtained uncertainties on the masses and the satellites' semi-major axis are given respectively in Table \ref{Res1Plu} and \ref{Res2Plu}.

As can be seen, Charon's mass would be considerably improved by those observations. Concerning Nix and Hydra's dynamics, the most interesting results concern their semi-major axis, whose uncertainty would be lowered even though they are not observed by GAIA. We can deduce that these observations, if they were to be available in time to prepare New Horizons' arrival, would be very precious to constrain not only Pluto's and Charon's motion, but those of Nix and Hydra as well. We can see here that in this case, strengthening the constraints on Pluto and Charon also constrains Nix and Hydra.

\begin{table}
\begin{center}

%\begin{tabular}{p{0.45\linewidth} p{0.46\linewidth}}
\begin{tabular}{p{0.45\linewidth} p{0.46\linewidth}}
\hline\hline 
 set of simulated observations & 1-$\sigma$ error bars on the masses (km$^3$s$^{-2}$)\\
 &number of simulated observations
\end{tabular}
\begin{tabular}{p{0.45\linewidth}p{0.2\linewidth}p{0.2\linewidth}}
% & \multicolumn{2}{c}{number of simulated observations}\\
  \hline
  & Pluto & Charon \\
\hline 
 1992-2014+NH & 0.25 & 0.045  \\ 
 	& & 181  \\
% 2002-2013+NH+GAIA & 0.089 & 0.0071 & 0.0017 & 0.0013 \\
% 	& 66 & 	247 & 158 & 176
 1992-2014+NH & 0.17 & 0.014  \\
  +GAIA	& & 181  \\
 1992-2015 & 0.45 & 0.035 \\
 & 41 & 166  \\
 
% (higher precision) & & & &\\
%    & 66 & 247 & 158 & 176\\
\hline 
& Nix & Hydra\\
\hline 
 1992-2014+NH & 0.0076 & 0.0026\\ 
 & 158 & 176\\ 
  1992-2014+NH& 0.0078 & 0.0024\\ 
  +GAIA & 186 & 233\\ 
   1992-2015  &0.0086 & 0.016\\ 
  & 68 & 69\\
  \hline
\end{tabular} 
\caption{ 1-$\sigma$ error bars on the masses given by least square method using different sets of simulated observations, using $m_P=870.3$ km$^3$ s$^{-2}$, $m_C=101.4$ km$^3$ s$^{-2}$, $m_N=0.039$ km$^3$ s$^{-2}$ and $m_H=0.021$ km$^3$ s$^{-2}$\cite{Tholen2008}.  }\label{Res1Plu}
\end{center}
\end{table}

\begin{table}
\begin{center}
\begin{tabular}{p{0.45\linewidth}p{0.46\linewidth}}
\hline\hline 
 set of simulated observations & 1-$\sigma$ error bars on the semi-major axis (km)
\end{tabular}
\begin{tabular}{p{0.35\linewidth}p{0.15\linewidth}p{0.15\linewidth}p{0.15\linewidth}}
  &  Charon & Nix & Hydra\\
  \hline 
 1992-2014 & 5.8 & 23 & 155\\
 1992-2015 & 3.25 & 10 & 43\\
\hline 
\end{tabular} 
\caption{1-$\sigma$ error bars on the semi-major axis using two different sets of simulations. The current estimation of the semi-major axis is $a_C=19570.45$ km, $a_N=49242$ km and $a_H=65082$ km \cite{Tholen2008}}\label{Res2Plu}
\end{center}
\end{table}

\section{Eugenia's system}
\subsection{Description}
(45)Eugenia is one of the few known triple asteroids. The primary, Eugenia, is far from spherical, its shape being obviously oblate from high resolution observations and light-curve inversion method. Its two satellites, Petit Prince and S2004(45) are quite close to their primary and far smaller than it. The outermost satellite, Petit Prince, is the first discovered satellite of the system \cite{Merline} in 1999, while the second one, closer to Eugenia, has been discovered in 2007 \cite{Franck2007}. Petit Prince semi-major axis is only 3\% of Eugenia's Hill radius, meaning both satellites are deep inside Eugenia's gravitational well. Yet, most recent studies of the system \cite{Franck2008,Franck2010} imply that they have a non-negligible inclination with respect to Eugenia's equatorial plane. From Eugenia's shape, a theoretical value of its second order polar oblateness has been estimated to be 0.19. This is in contradiction with a much lower value of 0.06 deduced from the satellite orbital motions. S2004(45) is always too close to Eugenia to be observed by GAIA considering the difference in their magnitude, about 8.2. On the contrary, Petit-Prince should be detectable when being close to its higher separation from Eugenia but the difference in their magnitude, about 7.4, prevents Petit Prince to be seen by GAIA when too close to Eugenia.
\begin{table}
%\begin{center}
\begin{tabular}{ccc}
\hline\hline &Eugenia&Petit Prince \\
\hline
semi-major axis & 2.720 AU &  1164.51 km \\
diameter& 214 km & ~13 km \\
angular diameter& 110 mas & 7 mas \\
magnitude & 7.46 & 16.8 \\
GM (km$^3$ s$^{-2}$)& 0.376 & $1.67\times 10^{-5}$\\
\hline
& S2004(45)\\
\hline
semi-major axis& 610.8 km\\
diameter& ~6 km\\
angular diameter& 3 mas\\
magnitude& -\\
GM (km$^3$ s$^{-2}$)&$1.67\times 10^{-5}$\\
\hline
\end{tabular}
\caption{Characteristics of Eugenia and its satellites}
\end{table}

\subsection{Data simulation}
We used again the simulation of asteroids'transit on GAIA's CCD with Rendez-vous software, and extracted the scheduled dates for Eugenia's observations.  The dates used are given in Table \ref{obseugsimgaia}. As Petit-Prince will not be seen when too close to Eugenia, we decided to reject the positions where he would be closer than 500 mas from Eugenia. We gave the observations a 1 mas uncertainty, as for Pluto's system.

\subsection{Results}

The uncertainties obtained for the semi-major axis of the satellites, and the pole orientation of Eugenia, are given in Table \ref{ResEug}. Concerning the second order polar oblateness, its uncertainty for the 1998-2010 set is 0.0006, and 0.0002 with GAIA observations, to be compared with $J_2=0.060$. As can be seen when comparing the two sets of observations, the orbital elements of Petit Prince will be constrained by GAIA observations, and as a result, the uncertainties on those of S2004(45) will also be lowered. We also find that the pole orientation will be more constrained since this data is obtained from the satellites'motion.

\begin{table}
%\begin{center}
\begin{tabular}{ccc}
\hline\hline 
semi major axis (km)&Petit Prince&S2004(45) \\
\hline
1998-2010 & 0.011 &  0.055  \\
1998-2010+GAIA & 0.005 & 0.029 \\
\hline
Eugenia's pole ($^{\circ}$)& $\lambda$ & $\beta$\\
\hline
1998-2010 & 0.20 & 0.10\\
1998-2010+GAIA & 0.040 & 0.041\\\hline
\end{tabular}
\caption{1-$\sigma$ error-bar on some of Eugenia's and its satellites' dynamical parameters using different sets of simulations}\label{ResEug}
\end{table}

\begin{table}
\caption{Dates used for the simulation of GAIA observations for Eugenia's system.}\label{obseugsimgaia}
\begin{center}
\begin{tabular}{cc}
\hline\hline
Dates  \\
\hline
01/09/2013 09:11:37.51  &  07/10/2015 02:52:09.03   \\
10/02/3013 02:44:17.81	&  07/10/2015 07:05:45.20   \\
10/02/3013 04:30:52.44	&  27/04/2016 06:37:21.13   \\
03/03/2013 20:42:14.56	&  27/04/2016 08:23:54.12   \\
03/03/2013 22:29:47.82	&  17/05/2016 02:26:15.30   \\
24/09/2013 08:31:35.33	&  20/06/2016 12:05:37.74   \\
24/09/2013 10:18:08.32	&  30/07/2016 05:24:55.81   \\
18/10/2013 02:25:42.47	&  30/07/2016 07:11:29.85   \\
18/10/2013 04:12:17.02	&  30/07/2016 01:25:09.47   \\
17/11/2013 02:02:57.09	&  30/07/2016 13:11:43.41   \\
17/11/2013 03:49:30.60	&  06/08/2016 11:30:38.04   \\
28/03/2014 22:32:12.19	&  06/08/2016 13:17:11.99   \\
28/03/2014 08:28:59.12	&  06/08/2016 17:30:51.61   \\
28/03/2014 10:15:33.75	&  06/08/2016 19:17:25.64   \\
22/05/2014 08:18:48.44	&  22/11/2016 01:34:33.66   \\
14/12/2014 03:38:45.72	&  23/12/2016 20:07:10.76   \\
14/12/2014 07:52:23.17	&  03/02/2017 17:31:58.57   \\
14/12/2014 09:38:56.17	&  03/02/2017 19:18:33.12   \\
26/12/2014 01:59:12.19	&  03/02/2017 23:32:13.78   \\
26/12/2014 03:45:46.66	&  04/02/2017 01:18:48.15   \\
03/02/2015 19:12:48.99	&  10/02/2017 23:37:53.85   \\
03/02/2015 20:59:21.72  &  11/02/2017 01:24:27.62   \\
07/03/2015 14:28:37.52	&  11/02/2017 05:38:06.55   \\
28/03/2015 06:40:05.29	&  11/02/2017 07:24:40.41   \\
27/06/2015 22:11:47.47	&  17/02/2017 12:16:15.89   \\
28/06/2015 02:25:26.05	&  17/02/2017 14:02:48.97   \\
28/06/2015 04:11:59.57	&  17/02/2017 18:16:27.21   \\
28/06/2015 08:25:37.98	&  30/07/2017 00:23:56.23   \\
28/06/2015 10:12:11.40	&  30/07/2017 02:10:30.95   \\
04/07/2015 04:16:12.89	&  30/07/2017 06:24:12.47   \\
04/07/2015 08:29:51.56	&  07/09/2017 11:37:45.64   \\
04/07/2015 10:16:25.25	&  07/09/2017 13:24:18.81   \\
04/07/2015 14:30:04.18	&  10/10/2017 17:08:09.17   \\
13/08/2015 15:26:02.11	&  10/10/2017 18:54:43.29   \\
14/09/2015 19:08:59.77	&  29/10/2017 05:12:32.26   \\
14/09/2015 20:55:34.32	&  29/10/2017 06:59:06.37   \\
\hline
\end{tabular}
\end{center}
\end{table}
\section{Conclusion}

For both systems, Pluto's and Eugenia's, GAIA will put new constrains on the dynamical properties of the system's bodies. 
Pluto has the advantage that it will be observed by both GAIA and New Horizons, and each mission will give us new highlight on the system.
The most interesting feature is that even though GAIA will not observe every body of the systems, the properties of the non-detected objects will also be better known than before the mission. The work presented here concerns only two multiple systems. Though the fainter components of each of them will not necessarily be detected, because of its absolute magnitude or relative magnitude to the primary and/or its closeness to the primary, GAIA will give us new informations about their dynamical behavior, masses and gravity fields, and hence new informations about the composition of the system.

\bibliographystyle{apalike}
\bibliography{bibPise}

\begin{thebibliography}{}

\bibitem[{Bancelin} et~al., 2011]{Bancelin}
{Bancelin}, D., {Hestroffer}, D., and {Thuillot}, W. (2011).
\newblock Dynamics of asteroids and neos from gaia astrometry.
\newblock {\em Planetary and Space Science}, this issue.

\bibitem[{Beauvalet} et~al., 2012]{Moi}
{Beauvalet}, L., {Lainey}, V., {Arlot}, J.-E., and {Binzel}, R.~P. (2012).
\newblock {Dynamical parameter determinations in Pluto's system. Expected
  constraints from the New Horizons mission to Pluto}.
\newblock {\em \aap}, 540:A65.

\bibitem[{Christy} and {Harrington}, 1978]{chardisc}
{Christy}, J.~W. and {Harrington}, R.~S. (1978).
\newblock {The satellite of Pluto}.
\newblock {\em Astron. J.}, 83:1005--+.

\bibitem[{Everhart}, 1985]{Everhart1985}
{Everhart}, E. (1985).
\newblock {An efficient integrator that uses Gauss-Radau spacings}.
\newblock In {A.~Carusi \& G.~B.~Valsecchi}, editor, {\em Dynamics of Comets:
  Their Origin and Evolution, Proceedings of IAU Colloq. 83, held in Rome,
  Italy, June 11-15, 1984. Edited by Andrea Carusi and Giovanni B. Valsecchi.
  Dordrecht: Reidel, Astrophysics and Space Science Library. Volume 115, 1985,,
  p.185}, pages 185--+.

\bibitem[{Marchis} et~al., 2007]{Franck2007}
{Marchis}, F., {Baek}, M., {Descamps}, P., {Berthier}, J., {Hestroffer}, D.,
  and {Vachier}, F. (2007).
\newblock {S/2004 (45) 1}.
\newblock {\em IAU Circular}, 8817:1--+.

\bibitem[{Marchis} et~al., 2008]{Franck2008}
{Marchis}, F., {Descamps}, P., {Baek}, M., {Harris}, A.~W., {Kaasalainen}, M.,
  {Berthier}, J., {Hestroffer}, D., and {Vachier}, F. (2008).
\newblock {Main belt binary asteroidal systems with circular mutual orbits}.
\newblock {\em Icarus}, 196:97--118.

\bibitem[{Marchis} et~al., 2010]{Franck2010}
{Marchis}, F., {Lainey}, V., {Descamps}, P., {Berthier}, J., {van Dam}, M., {de
  Pater}, I., {Macomber}, B., {Baek}, M., {Le Mignant}, D., {Hammel}, H.~B.,
  {Showalter}, M., and {Vachier}, F. (2010).
\newblock {A dynamical solution of the triple asteroid system (45) Eugenia}.
\newblock {\em Icarus}, 210:635--643.

\bibitem[{Merline} et~al., 1999]{Merline}
{Merline}, W.~J., {Close}, L.~M., {Dumas}, C., {Chapman}, C.~R., {Roddier}, F.,
  {Menard}, F., {Slater}, D.~C., {Duvert}, G., {Shelton}, C., and {Morgan}, T.
  (1999).
\newblock {Discovery of a moon orbiting the asteroid 45 Eugenia}.
\newblock {\em Nature}, 401:565--+.

\bibitem[{Showalter} et~al., 2011]{2011P4}
{Showalter}, M.~R., {Hamilton}, D.~P., {Stern}, S.~A., {Weaver}, H.~A.,
  {Steffl}, A.~J., and {Young}, L.~A. (2011).
\newblock {New Satellite of (134340) Pluto: S/2011 (134340) 1}.
\newblock {\em IAU Circular}, 9221:1.

\bibitem[{Standish}, 1998]{Standish1998}
{Standish}, E.~M. (1998).
\newblock Technical report, Jet Prop. Lab. Interoffice Memo. 312.F-98-04.

\bibitem[{Stern} et~al., 2006]{2006astro.ph..5014S}
{Stern}, S.~A., {Mutchler}, M.~J., {Weaver}, H.~A., and {Steffl}, A.~J. (2006).
\newblock {The Positions, Colors, and Photometric Variability of Pluto's Small
  Satellites from HST Observations 2005-2006}.
\newblock {\em ArXiv Astrophysics e-prints}.

\bibitem[{Tholen} et~al., 2008]{Tholen2008}
{Tholen}, D.~J., {Buie}, M.~W., {Grundy}, W.~M., and {Elliott}, G.~T. (2008).
\newblock {Masses of Nix and Hydra}.
\newblock {\em Astronomical Journal}, 135:777--784.

\bibitem[{Weaver} et~al., 2005]{Weaver2005}
{Weaver}, H.~A., {Stern}, S.~A., {Mutchler}, M.~J., {Steffl}, A.~J., {Buie},
  M.~W., {Merline}, W.~J., {Spencer}, J.~R., {Young}, E.~F., and {Young}, L.~A.
  (2005).
\newblock {S/2005 P 1 and S/2005 P 2}.
\newblock {\em IAU Circ.}, 8625:1--+.

\end{thebibliography}

%% not want to use model2-names.bst.



%% \bibitem must have one of the following forms:
%%   \bibitem[Jones et al.(1990)]{key}...
%%   \bibitem[Jones et al.(1990)Jones, Baker, and Williams]{key}...
%%   \bibitem[Jones et al., 1990]{key}...
%%   \bibitem[\protect\citeauthoryear{Jones, Baker, and Williams}{Jones
%%       et al.}{1990}]{key}...
%%   \bibitem[\protect\citeauthoryear{Jones et al.}{1990}]{key}...
%%   \bibitem[\protect\astroncite{Jones et al.}{1990}]{key}...
%%   \bibitem[\protect\citename{Jones et al., }1990]{key}...
%%   \harvarditem[Jones et al.]{Jones, Baker, and Williams}{1990}{key}...
%%


% \end{thebibliography}

\end{document}